\documentclass[final]{aipproc}
\layoutstyle{8x11double}
\usepackage{amsfonts}
\usepackage{amsmath}
\usepackage{bm}
\usepackage{amssymb}
\usepackage{graphicx}
\begin{document}

\title{Control of the chaotic velocity dispersion of a cold electron beam interacting with electrostatic waves}
\author{G. Ciraolo} {address={Association Euratom-CEA, DRFC/DSM/CEA, CEA Cadarache,
F-13108 St. Paul-lez-Durance Cedex, France}}
\author{C. Chandre}{address={Centre de Physique Th\'eorique, CNRS Luminy, Case 907, F-13288 Marseille Cedex 9,
France}\thanks{Unit\'e Mixte de Recherche (UMR 6207) du CNRS, et des
universit\'es Aix-Marseille I, Aix-Marseille II et du Sud
Toulon-Var. Laboratoire affili\'e \`a la FRUMAM (FR 2291).}}
\author{R. Lima}{address={Centre de Physique Th\'eorique, CNRS Luminy, Case 907, F-13288 Marseille Cedex 9,
France}\thanks{Unit\'e Mixte de Recherche (UMR 6207) du CNRS, et des
universit\'es Aix-Marseille I, Aix-Marseille II et du Sud
Toulon-Var. Laboratoire affili\'e \`a la FRUMAM (FR 2291).}}
\author{M. Pettini}{address={Istituto Nazionale di Astrofisica, Largo Enrico Fermi 5, I-50125 Firenze, Italy,
INFM UdR Firenze and INFN Sezione di Firenze}}
\author{M. Vittot}{address={Centre de Physique Th\'eorique, CNRS Luminy, Case 907, F-13288 Marseille Cedex 9,
France}\thanks{Unit\'e Mixte de Recherche (UMR 6207) du CNRS, et des
universit\'es Aix-Marseille I, Aix-Marseille II et du Sud
Toulon-Var. Laboratoire affili\'e \`a la FRUMAM (FR 2291).}}

\begin{abstract}
In this article we present an application of a method of control of
Hamiltonian systems to the chaotic velocity diffusion of a cold
electron beam interacting with electrostatic waves. We numerically
show the efficiency and robustness of the additional small control
term in restoring kinetic coherence of the injected electron beam.
\end{abstract}
%\begin{keyword}
%Hamiltonian systems \sep control
%\PACS \\
%05.45.-a Nonlinear dynamics and nonlinear dynamical systems \\
%05.45.Gg Control of chaos, applications of chaos
%\end{keyword}
\maketitle

%===========================================================
\section{Introduction}
%===========================================================
The consequences of chaotic dynamics can be harmful in several
contexts. During the last decade or so, much attention has been paid
to the problem of chaos control. Controlling chaos means that one
aims at reducing or suppressing chaos by mean of a small
perturbation so that the original structure of the system under
investigation is kept practically unaltered, while its behavior can
be substantially altered. In particular, in many physical devices
there are undesirable effects due to transport phenomena that can be
attributed to a chaotic dynamics. For example, chaos in beams of
particle accelerators leads to a weakening of the beam
luminosity~\cite{acceleratori,robin}. Similar problems are
encountered in free electron lasers~\cite{boni90,deni04}. In
magnetically confined fusion plasmas, the so called anomalous
transport, which has its microscopic origin in a chaotic transport
of charged particles, represents a challenge to the attainment of
high performance in fusion devices~\cite{carre97}. The theoretical
description of the physics of these devices (beam accelerators, free
electron lasers, fusion devices) is based on Hamiltonian models,
thus the above mentioned possibility of harmful consequences of
chaos is related with Hamiltonian chaos. One way to control
transport would be that of reducing or suppressing chaos. There
exist numerous attempts to cope with this problem~\cite{rev_c_d}.
However, these efforts have mainly been focused on dissipative
systems. For Hamiltonian systems, the absence of phase space
attractors has been a hindrance to the development of efficient
methods of control. Conventional approaches aim at targeting
individual trajectories. However, when a microscopic description of
a physical system is required, targeting of individual microscopic
trajectories is hopeless.

The strategy we developed is based on a small but suitable
modification of the original Hamiltonian system such that the
controlled one has a more regular dynamics. In this article this
strategy is illustrated through a particular example: how to restore
kinetic coherence of a cold electron beam interacting with
electrostatic waves. Here chaos deteriorates the velocity
"monochromaticity" of the injected beam and one has to find a
strategy of control such that this effect is reduced or suppressed.

The Hamiltonian which models the dynamics of charged test particles
moving in electrostatic waves is
\begin{equation}
H(p,x,t)=\frac{p^2}{2}+\sum_{n=1}^N\varepsilon_n\cos
(k_nx-\omega_nt+\varphi_n), \label{Hstart}
\end{equation}
where $p$ is the test charged particle momentum and $\varepsilon_n$,
$k_n$, $\omega_n$ and $\varphi_n$ are respectively the amplitudes,
wave numbers, frequencies  and phases of the electrostatic waves.
Generically, for $N\geq 2$, the dynamics of the particles governed
by this equation is a mixture of regular and chaotic behaviours
mainly depending on the amplitude of the waves.
\begin{figure}
\centering
\includegraphics[width=\textwidth]{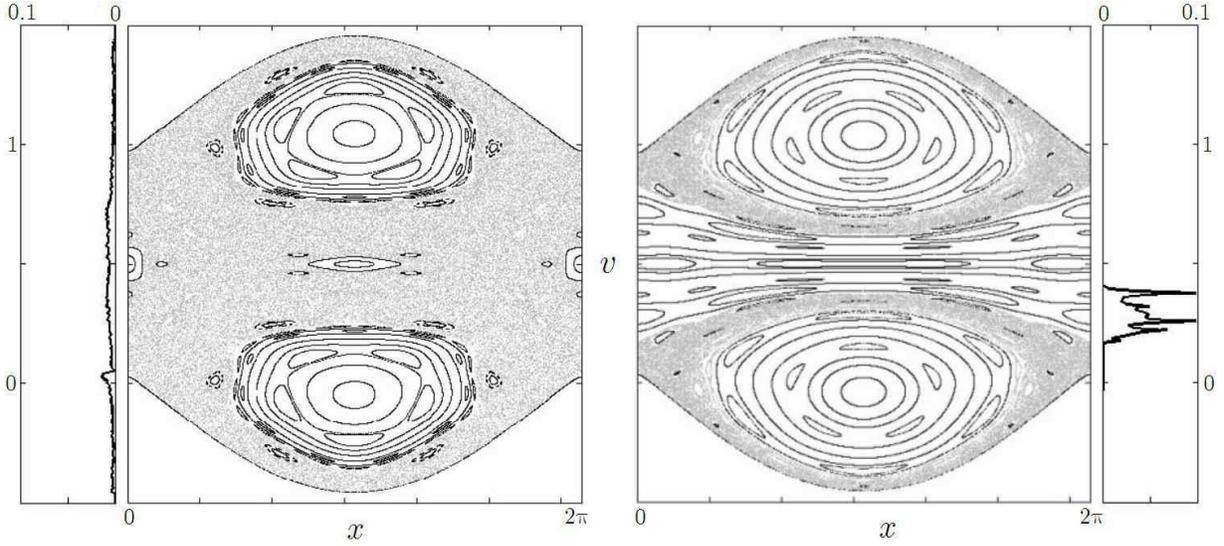}
\caption{Poincar\'e surface of section and (normalized) probability
distribution functions $F(p)$ of the velocity for the Hamiltonian
given by Eq.~(\ref{Hstart}) without control term (left panel) and
with the control term given by Eq.~(\ref{eqn:f2app}) (right panel).
The depicted numerical results are obtained in the case of two
electrostatic waves using
$(\varepsilon_1,k_1,\omega_1,\varphi_1)=(0.045,1,0,0)$ and
$(\varepsilon_2,k_2,\omega_2,\varphi_2)=(0.045,1,1,0)$. Reprinted
from Ref.~\cite{TWT05}.}\label{fig_num_twt}
\end{figure}
A Poincar\'e section of the dynamics (a stroboscopic plot of
selected trajectories) is depicted on Fig.~\ref{fig_num_twt} for two
waves ($N=2$). A large zone of chaotic behaviour occurs in between
the primary resonances (nested regular structures). If one considers
a beam of initially monokinetic particles, this zone is associated
with a large spread of the velocities after some time (left hand
side panel of Fig.~\ref{fig_num_twt}) since the particles are moving
in the chaotic sea created by the overlap of the two
resonances~\cite{chir79}. Here the problem of control is to find a
control term  $f$ given by
\begin{equation}
f(x,t)=\sum_{n=1}^q\eta_n\cos(\kappa_nx-\Omega_nt+\Phi_n),
\end{equation}
that is the right set of additional controlling waves, such that the
controlled Hamiltonian
$$
H_c(p,x,t)=H(p,x,t)+f(x,t),
$$
given by the original one plus the control term, is more regular
than $H$. In general, adding a generic perturbation adds more
resonances and hence more chaos. Nevertheless this problem has some
obvious solutions with additional waves having amplitudes of the
same order as the initial ones. For energetic purposes, the
additional waves must have amplitudes which are much smaller than
the initial ones
($\max_{n}\vert\eta_n\vert\ll\max_{n}\vert\varepsilon_n\vert$). This
means finding a specifically designed small control term that is a
slight modification of the system but which drastically changes the
dynamical behaviour from chaotic to regular. In particular one aims
at building barriers to transport that prevent large scale chaos to
occur in the system.

\section{Local control method and control term}
In this section we briefly sketch the local control method that was
extensively discussed in Refs.~\cite{vitt05,chan05}. Let us consider
Hamiltonian systems of the form
$$
H({\bf A},{\bm\theta})=H_0({\bf A})+\varepsilon V({\bf
A},{\bm\theta}),
$$
that after a suitable expansion can be rewritten
as
\begin{equation}
\label{eqn:ham}
    H({\bf A},{\bm\theta})={\bm \omega}\cdot {\bf A}+ \varepsilon v({\bm \theta}) +w({\bf A},{\bm\theta}),
\end{equation}
where $({\bf A},{\bm\theta})\in {\mathbb R}^L\times {\mathbb T}^L$
are action-angle variables for $H_0$ in a phase space of dimension
$2L$. The last term of the expansion can be written as
$$
w({\bf A},{\bm\theta})=\varepsilon {\bf w}_1({\bm\theta})\cdot {\bf
A}+ w_2({\bf A},{\bm\theta}),
$$
where $w_2$ is quadratic in the actions, i.e.\ $w_2({\bf
0},{\bm\theta})=0$ and $\partial_{\bf A}w_2({\bf
0},{\bm\theta})={\bf 0}$. We assume that $\bm\omega$ is a
non-resonant vector of ${\mathbb R}^L$, i.e.\ there is no non-zero
${\bf k}\in {\mathbb Z}^L$ such that ${\bm\omega}\cdot {\bf k}=0$.
Moreover we assume without restrictions that $\int_{{\mathbb T}^L}
v({\bm\theta})d^L{\bm\theta}=0$.

We consider a region near ${\bf A}={\bf 0}$. The controlled
Hamiltonian we construct is given by
\begin{equation}
\label{eqn:gene} H_c({\bf A},{\bm\theta})={\bm \omega}\cdot {\bf
A}+\varepsilon v({\bm \theta}) +w({\bf A},{\bm\theta}) +
\varepsilon^2 f({\bm \theta}).
\end{equation}
The control term depends only on angle variables and its expression
is given by
\begin{equation}
\label{eqn:CT}
    f({\bm\theta})=-\varepsilon^{-2} w(-\varepsilon \Gamma \partial_{\bm\theta} v,{\bm\theta}),
\end{equation}
where $\partial_{\bm\theta}v$ denotes the first derivatives of $v$
with respect to ${\bm\theta}$~: $ \partial_{\bm\theta}v=\sum_{{\bf
k}\in{\mathbb Z}^L}i{\bf k} v_{\bf k} {\mathrm e}^{i{\bf k}\cdot
{\bm\theta}}$ and where the linear operator $\Gamma$ is a
pseudo-inverse of ${\bm \omega}\cdot\partial_{\bm\theta}$, i.e. its
action on a function $v({\bm\theta})=\sum_{{\bf k}\in{\mathbb
Z}^L}v_{\bf k} {\mathrm e}^{i{\bf k}\cdot {\bm\theta}}$ is~:
\begin{equation}
\label{op_gamma} \Gamma v({\bm\theta})=\sum_{{\bf k}\in{\mathbb
Z}^L\setminus \{ {\bf 0}\} } \frac{v_{\bf k}} {i{\bm \omega}\cdot
{\bf k}} {\mathrm e}^{i{\bf k}\cdot {\bm\theta}}.
\end{equation}

We prove (see Ref.~\cite{chan05}) that $H_c$ has an invariant torus
located at ${\bf A}=-\varepsilon \Gamma \partial_{\bm\theta} v$. For
Hamiltonian systems with two degrees of freedom, such an invariant
torus acts as a barrier to diffusion. For the construction of the
control term, we notice that we do not require that the quadratic
part of $w$ is small in order to have a control term of order
$\varepsilon^2$.

\subsection{Control term for a two wave model}
We consider the following Hamiltonian with two traveling waves:
\begin{eqnarray}
\label{eqn:Hfp}
    H(p,x,t)&=&\frac{p^2}{2}+\varepsilon_1\cos (k_1 x-\omega_1 t+\varphi_1)\nonumber\\
    && +\varepsilon_2\cos (k_2 x-\omega_2 t+\varphi_2),
\end{eqnarray}
where the wavenumbers are chosen according to a given dispersion
relation $k_1=K(\omega_1)$ and $k_2=K(\omega_2)$. The numerical
results we will show in the following have been obtained using
$(\varepsilon_1,k_1,\omega_1,\varphi_1)=(\varepsilon,1,0,0)$ and
$(\varepsilon_2,k_2,\omega_2,\varphi_2)=(\varepsilon,1,1,0)$.
Similar results are expected for other values of
$(\varepsilon_n,k_n,\omega_n,\varphi_n)$.

\begin{figure}
\centering
\includegraphics[width=0.4\textwidth]{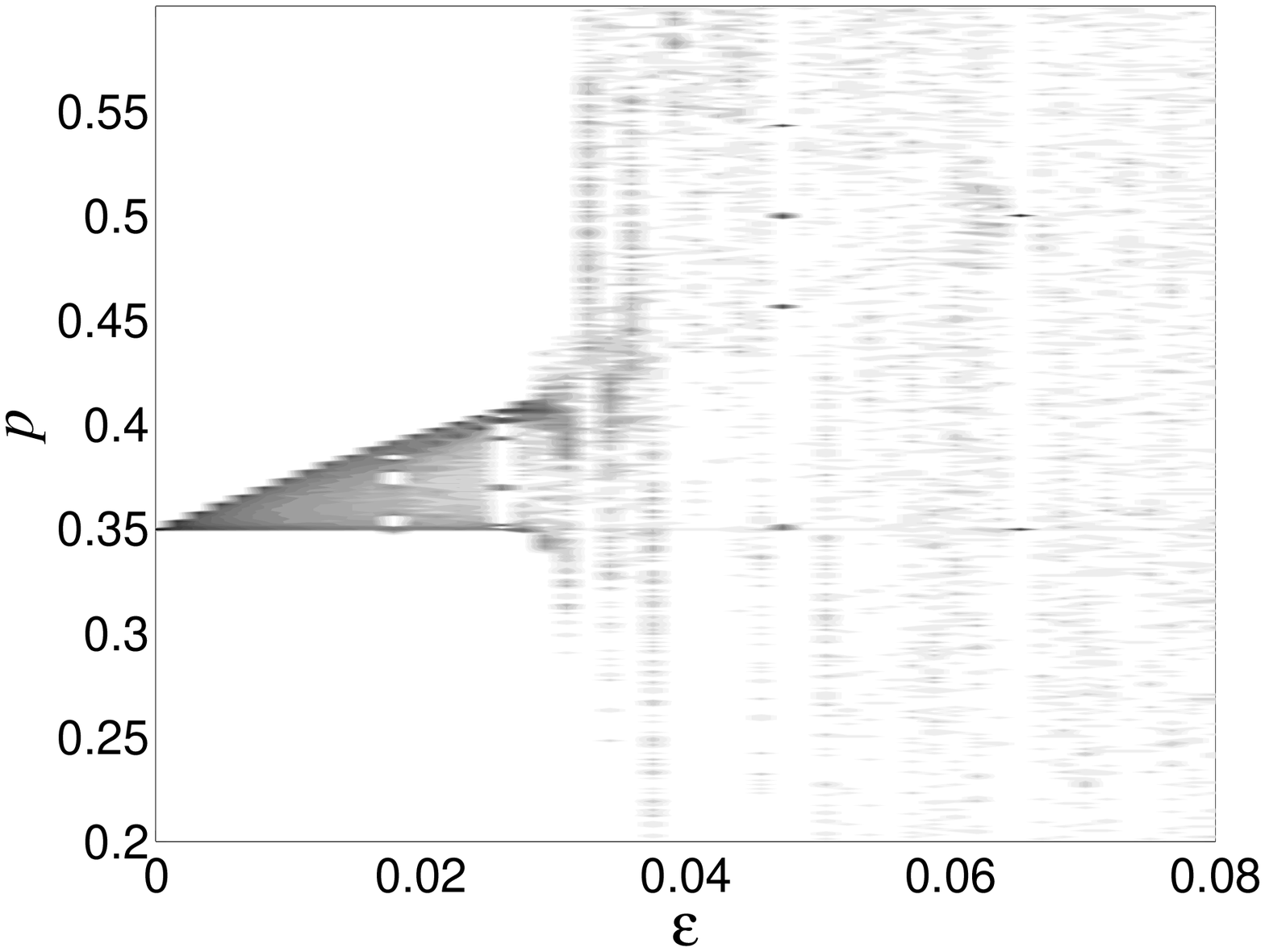}
\includegraphics[width=0.4\textwidth]{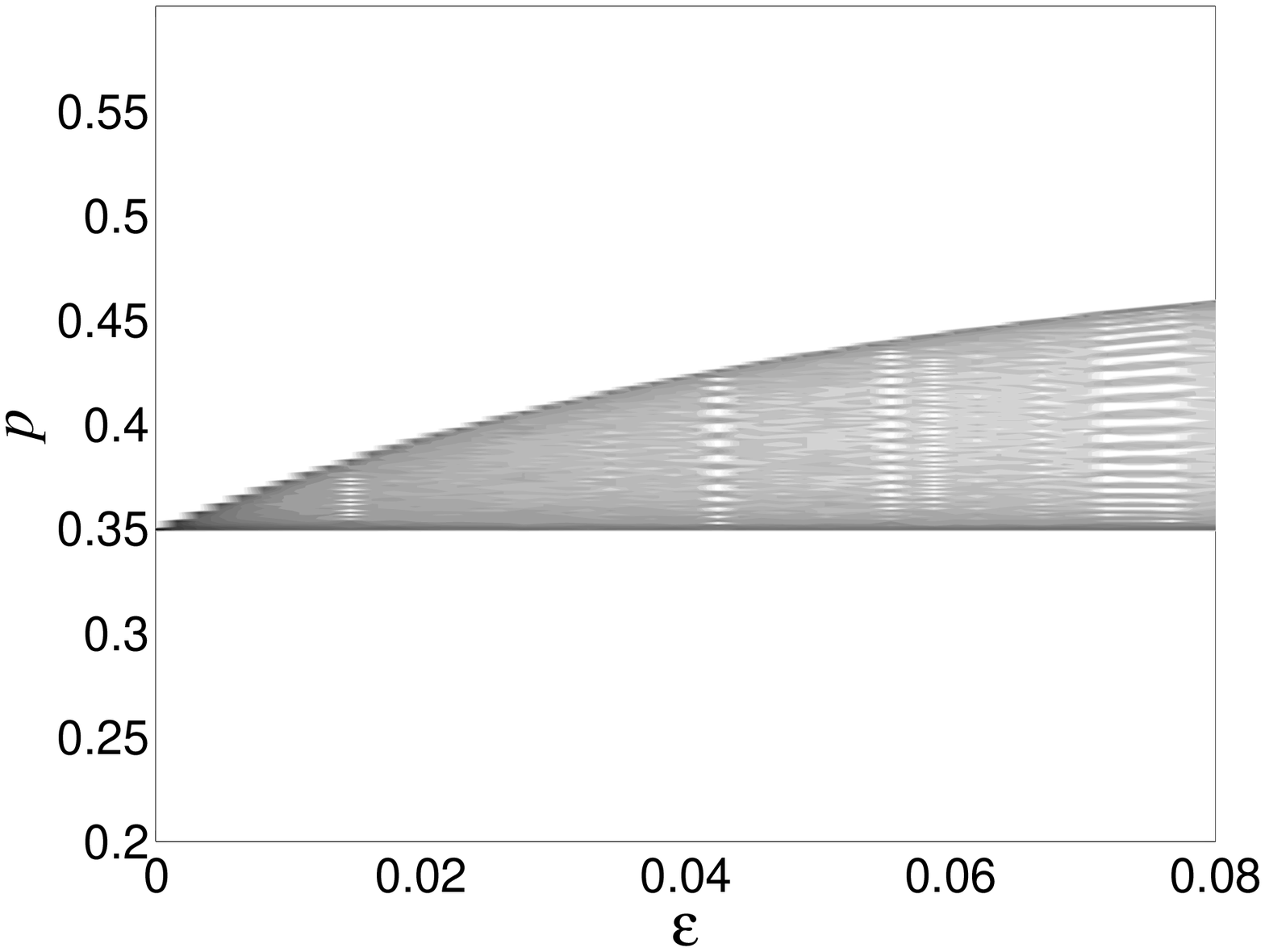}
\caption{Distribution of momenta $p$ as a function of $\varepsilon$
for Hamiltonian~(\ref{eqn:Hfp}) (left panel) and for
Hamiltonian~(\ref{eqn:Hfp}) plus the control term~(\ref{eqn:fTWT})
(right panel). The numerical results are obtained for the following
values of the parameters:
$(\varepsilon_1,k_1,\omega_1,\varphi_1)=(\varepsilon,1,0,0)$ and
$(\varepsilon_2,k_2,\omega_2,\varphi_2)=(\varepsilon,1,1,0)$.}
\label{figure1TWT}
\end{figure}
Figure~\ref{figure1TWT} (left panel) depicts the probability
distribution function of the momenta of a trajectory of
Hamiltonian~(\ref{eqn:Hfp}) for $t\in[0,800\pi]$ (with initial
condition chosen in the chaotic sea, e.g. for $(p\approx 0.35,
x=0)$) as a function of the amplitude $\varepsilon$ of the
perturbation. It shows that after $\varepsilon_c\approx 0.028$ there
is no longer any barrier in phase space. We notice that the value of
$\varepsilon_c$ for which the last invariant torus is broken is
approximately equal to $\varepsilon_c\approx 0.02759$~\cite{chan02}.

The first step in order to compute the local control term which
recreates a barrier in phase space is to map this Hamiltonian with
one and a half degrees of freedom into an autonomous Hamiltonian
with two degrees of freedom by considering that $t \mbox{ mod }2\pi$
is an additional angle variable. We denote $E$ its conjugate action.
The autonomous Hamiltonian is
\begin{eqnarray}
\label{Htwt_autonoma} H&=&E+\frac{p^2}{2}+\varepsilon_1\cos
(k_1x-\omega_1 t+\varphi_1)\nonumber\\
&&+\varepsilon_2\cos (k_2 x-\omega_2 t+\varphi_2).
\end{eqnarray}
Then, the momentum $p$ is shifted by $\omega$ in order to define a
local control in the region $p\approx 0$. The Hamiltonian is
rewritten as
\begin{eqnarray}
\label{Htwt_pshift} H&=&E+\omega p+\varepsilon_1\cos (k_1 x-\omega_1
t+\varphi_1)\nonumber\\
&&+\varepsilon_2\cos (k_2 x-\omega_2 t+\varphi_2)+\frac{p^2}{2}.
\end{eqnarray}
We define
$$
H_0=E+\omega p,
$$
$$
v(x,t)=\varepsilon_1\cos (k_1 x-\omega_1 t+\varphi_1) +\varepsilon_2\cos (k_2 x-\omega_2 t+\varphi_2),
$$

$$
w(p,x,t)=p^2/2,
$$
provided $\omega\neq\omega_1/k_1$ and  $\omega\neq\omega_2/k_2$.
In this case the control term is given by
$$
f(x,t)=-\frac{1}{2}(\Gamma \partial_{x} v) ^2,
$$
where $\partial_x$ is the partial derivative with respect to $x$ and
the action of $\Gamma$ follows from Eq.~(\ref{op_gamma}). Therefore
the control term is equal to
\begin{eqnarray}
    f(x,t)&=&-\frac{1}{2}\left[ \varepsilon_1k_1\frac{\cos (k_1x-\omega_1t+\varphi_1)}{\omega k_1-\omega_1}\right.\nonumber\\
    &&+\left.\varepsilon_2k_2\frac{\cos(k_2x-\omega_2t+\varphi_2)}{\omega k_2-\omega_2}\right]^2.
    \label{eqn:fTWT}
\end{eqnarray}

Adding the exact control term (\ref{eqn:fTWT}) to the
Hamiltonian~(\ref{Htwt_pshift}), an invariant KAM torus with
frequency $\omega$ is recreated. This barrier prevents the electron
beam to diffuse in phase space and the electron kinetic coherence is
restored. In Fig.~\ref{figure1TWT} (right panel) the recreation of
barrier in phase space can be observed on the probability
distribution function of the momenta of a trajectory of
Hamiltonian~(\ref{eqn:Hfp}) plus the control term~(\ref{eqn:fTWT})
for $t\in[0,800\pi]$ (with initial condition chosen as in the
uncontrolled case, e.g. for $(p\approx 0.35, x=0)$). In fact for all
the values of the amplitude $\varepsilon$ of the perturbation, the
kinetic coherence of the initial electron beam is preserved.

The local control ensures that the barrier persists for all the
magnitudes of the perturbation and also gives us explicitly the
equation of the recreated KAM torus, that is
\begin{eqnarray}
p_0(x,t)&=&\omega-\varepsilon_1k_1\frac{\cos
(k_1x-\omega_1t+\varphi_1)}{\omega
k_1-\omega_1}\nonumber\\
&&-\varepsilon_2k_2\frac{\cos(k_2x-\omega_2t+\varphi_2)}{\omega
k_2-\omega_2}.
\end{eqnarray}

The control term~(\ref{eqn:fTWT}) has four Fourier modes,
$(2k_1,-2\omega_1)$,  $(2k_2,-2\omega_2)$,
$((k_1+k_2),-(\omega_1+\omega_2))$ and
$((k_1-k_2),-(\omega_1-\omega_2))$. If we consider $\omega$ close to
$(\omega_1/k_1+\omega_2/k_2)/2$, the main Fourier mode of the
control term is
\begin{eqnarray}
\label{eqn:f2app}
f_2&=&-\frac{\varepsilon_1\varepsilon_2k_1k_2}{2(\omega
k_1-\omega_1)(\omega k_2-\omega_2)}\times\nonumber\\
&&\cos [(k_1+k_2)x-(\omega_1+\omega_2)t+\varphi_1+\varphi_2].~~
\end{eqnarray}

Assuming $\omega=(\omega_1/k_1+\omega_2/k_2)/2$, the control term given by Eq.~(\ref{eqn:f2app}) is
$$
f_2=\frac{2\varepsilon_1\varepsilon_2}{(v_1-v_2)^2}\cos
[(k_1+k_2)x-(\omega_1+\omega_2)t+\varphi_1+\varphi_2],
$$
where $v_n$ is the group velocity of the wave $n$ defined as
$v_n=\omega_n/k_n$.

\begin{figure}
\centering
\includegraphics[width=0.4\textwidth]{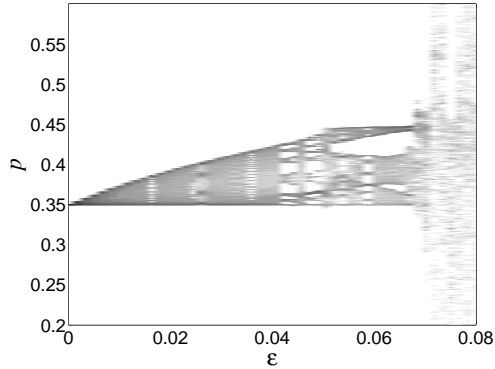}
\caption{Distribution of momentum $p$ as a function of $\varepsilon$
for Hamiltonian~(\ref{eqn:Hfp}) with the approximate control
term~(\ref{eqn:f2app}). The numerical results are obtained for the
same values of the parameters as in Fig.~\ref{figure1TWT}.}
\label{figure2TWT}
\end{figure}
Figure~\ref{figure2TWT} depicts the probability distribution
function of the momenta of a trajectory of
Hamiltonian~(\ref{eqn:Hfp}) with the approximate control
term~(\ref{eqn:f2app}), for $t\in[0,800\pi]$ and with initial
condition $(p\approx 0.35, x=0)$, as a function of the amplitude
$\varepsilon$ of the perturbation. It shows that even after
$\varepsilon_c\approx 0.028$ barriers in phase space have been
created. The approximate control is efficient till
$\varepsilon_c\approx 0.07$.

\subsection{Robustness of the control: Influence of a change of
frequency and phase in the control term} \label{sec:4}

\begin{figure}
\centering
\includegraphics[width=0.4\textwidth]{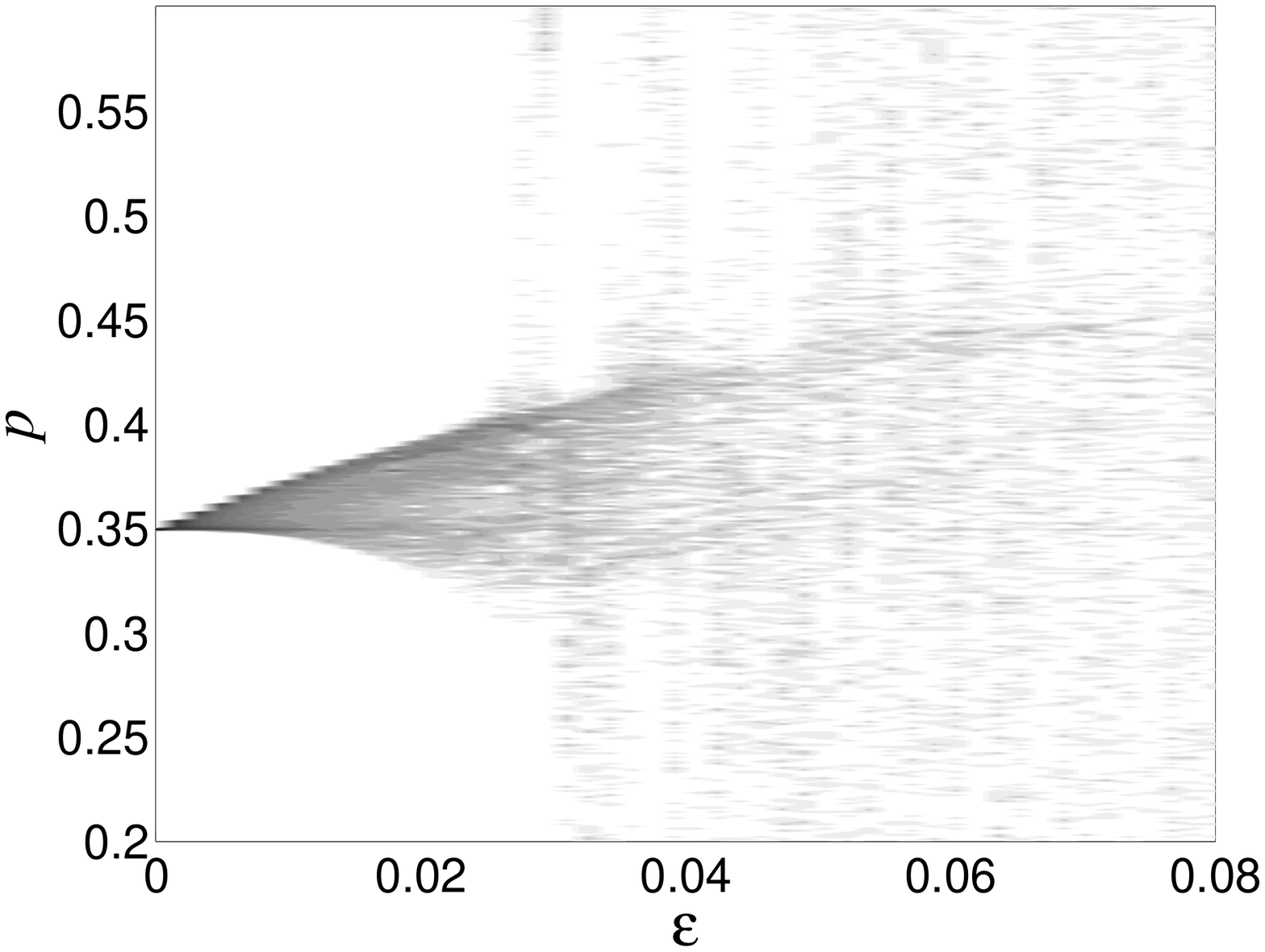}
\includegraphics[width=0.4\textwidth]{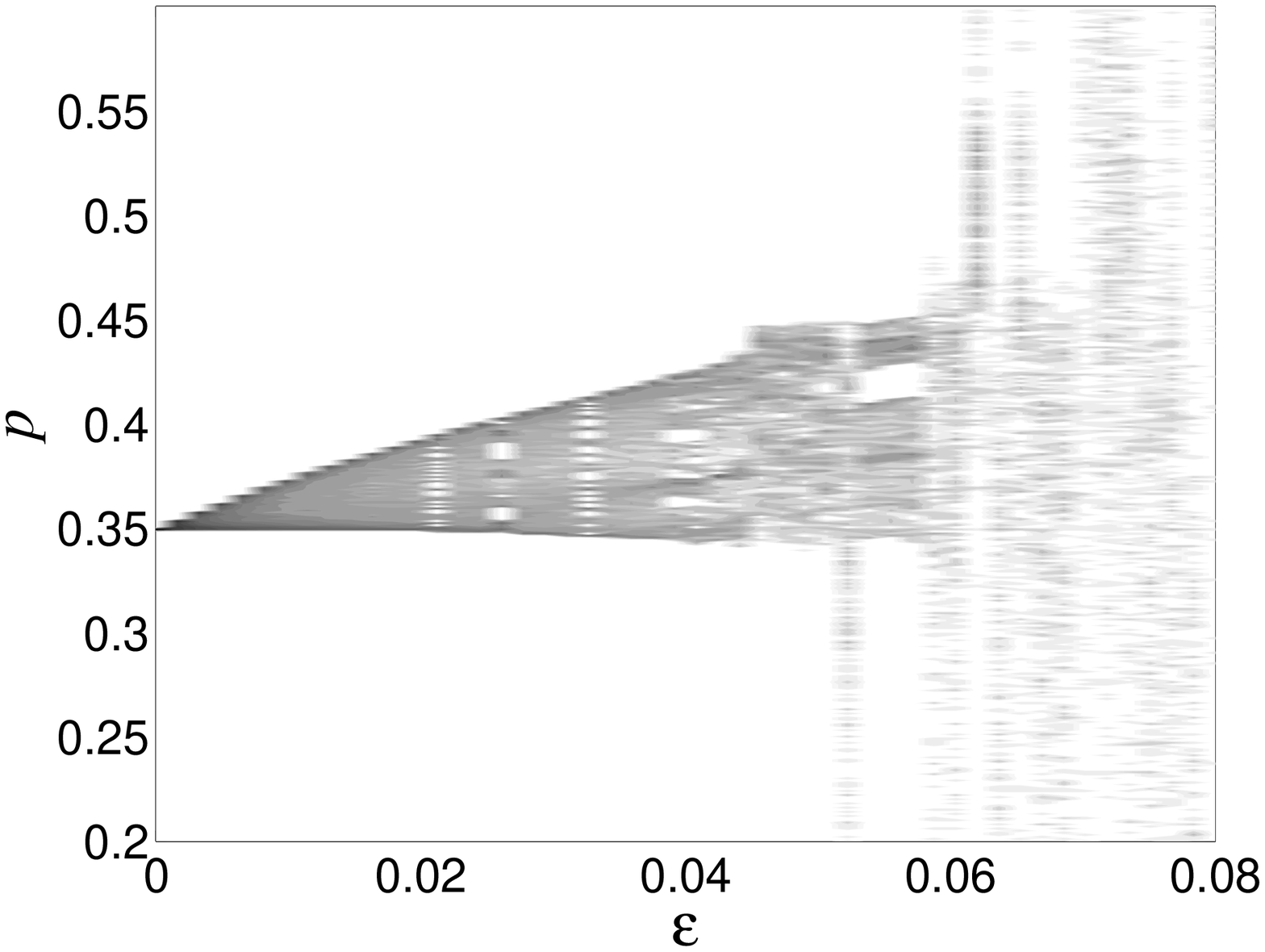}
\caption{Distribution of momentum $p$ as a function of $\varepsilon$
for Hamiltonian~(\ref{eqn:Hfp}) with the approximate control
term~(\ref{eqn:f2app}) containing an error of $\delta\omega=0.001$
in the frequency (left panel) and  an error of $\delta \varphi=0.2$
in the phase (right panel). The numerical results are obtained for
the same values of the parameters as in Fig.~\ref{figure1TWT}.}
\label{figure3TWT}
\end{figure}

The main problem is that the wavenumber $k_1+k_2$ of the control
term does not satisfy in general the dispersion relation
$k=K(\omega)$, i.e., we do not have in general
$k_1+k_2=K(\omega_1+\omega_2)$ since the dispersion relation is
not linear. Therefore the determination of the frequency and
wavenumber of the control term contains errors. The approximate
control term has the form
\begin{eqnarray}
f_2^{(\omega)}&=&-\frac{\varepsilon_1\varepsilon_2k_1k_2}{2(\omega
k_1-\omega_1)(\omega k_2-\omega_2)}\times\nonumber\\
&&\cos[(k_1+k_2)x-(\omega_1+\omega_2+\delta\omega)t+\varphi_1+\varphi_2].\nonumber
\end{eqnarray}
We also test the robustness of the local control when there is an
error on the phase, i.e. we consider an approximate control term of
the form
\begin{eqnarray}
f_2^{(\varphi)}&=&-\frac{\varepsilon_1\varepsilon_2k_1k_2}{2(\omega
k_1-\omega_1)(\omega k_2-\omega_2)}\times\nonumber\\
&&\cos[(k_1+k_2)x-(\omega_1+\omega_2)t+\varphi_1+\varphi_2+\delta\varphi].\nonumber
\end{eqnarray}
Figure~\ref{figure3TWT} depicts the probability
distribution function of the momenta of a trajectory of
Hamiltonian~(\ref{eqn:Hfp}) with the approximate control term
$f_2^{(\omega)}$ (left panel), or with $f_2^{(\varphi)}$ (right
panel), for $t\in[0,800\pi]$ and with initial condition $(p\approx
0.35, x=0)$, as a function of the amplitude $\varepsilon$ of the
perturbation. From these figures we notice that it is important to
adjust precisely the frequency of the control term, and also choose
a set of modes such that $k_1+k_2$ is close to
$K(\omega_1+\omega_2)$.

In summary, we have shown here how our method of control works in
suppressing chaotic velocity diffusion induced in a cold electron
beam interacting with electrostatic waves. These numerical results
have been successfully confirmed by an experimental check performed
on a Traveling Wave Tube~\cite{TWT05}. According to the fact that if
the amplitude of the potential is of order $\varepsilon$ then the
amplitude of the control term is of order $\varepsilon^2$, the
control term is realized with an additional cost of energy that is
less than $0.1\%$ of the initial electrostatic energy of the two
wave system.

%===========================================================
\section{Acknowledgements}
We acknowledge useful discussions with F. Doveil, Y. Elskens, Ph.
Ghendrih and A. Macor. We acknowledge the financial support from
Euratom/CEA (contract EUR 344-88-1 FUA F).
%===========================================================
%

\end{document}